\pdfoutput=1

\documentclass{article}
\usepackage{spconf,amsmath,graphicx}
\usepackage{amsfonts,mathrsfs}
\usepackage{bm}

\title{phoneme-based distribution regularization for SPEECH ENHANCEMENT }
\name{Yajing Liu\textsuperscript{1}\sthanks{This work was done when Yajing Liu was an intern at MSRA.}, Xiulian Peng\textsuperscript{2}, Zhiwei Xiong\textsuperscript{1}, Yan Lu\textsuperscript{2}}
\address{\textsuperscript{1}University of Science and Technology of China, Hefei, China\\
         \textsuperscript{2}Microsoft Research Asia, Beijing, China}
\begin{document}
%
\topmargin=0mm
\maketitle
\begin{abstract}
Existing speech enhancement methods mainly separate speech from noises at the signal level or in the time-frequency domain. They seldom pay attention to the semantic information of a corrupted signal. In this paper, we aim to bridge this gap by extracting phoneme identities to help speech enhancement. Specifically, we propose a phoneme-based distribution regularization (PbDr) for speech enhancement, which incorporates frame-wise phoneme information into speech enhancement network in a conditional manner. As different phonemes always lead to different feature distributions in frequency, we propose to learn a parameter pair, i.e. scale and bias, through a phoneme classification vector to modulate the speech enhancement network. The modulation parameter pair includes not only frame-wise but also frequency-wise conditions, which effectively map features to phoneme-related distributions. In this way, we explicitly regularize speech enhancement features by recognition vectors. Experiments on public datasets demonstrate that the proposed PbDr module can not only boost the perceptual quality for speech enhancement but also the recognition accuracy of an ASR system on the enhanced speech. This PbDr module could be readily incorporated into other speech enhancement networks as well.

\end{abstract}
 \vspace{-0.25cm}
\section{Introduction}
\label{sec:intro}
The quality and intelligibility of speech are deteriorated by noises in real world, which has a negative effect on speech communication and recognition \cite{wang2016joint,benesty2005speech}. To address this problem, speech enhancement is usually performed to separate the clean speech from background noises with reverberation. Thanks to the advancement of computational auditory scene analysis\cite{wang2006computational,pirhosseinloo2017time}, speech enhancement has been viewed as a supervised learning problem. Existing speech enhancement methods mainly learn a deep-learning-based neural network, which can effectively model the relationship between a noise-corrupted signal and the clean speech.
However, they mostly focus on the signal-level or time-frequency domain separation, neglecting the semantic information of the corrupted speech. In this paper, we aim to utilize speech recognition information to help speech enhancement network so that it could achieve better perceptual quality. 

There are several studies along this line, which introduce phoneme information to help speech enhancement by many phoneme-specific networks or simply concatenating known phoneme input with the noisy signal \cite{wang2016phoneme,schulze2020joint,chazan2016phoneme}. However, some are too complex as each phoneme corresponds to one enhancement network and some need known phoneme as an additional input. Incorrectly predicted phoneme labels will lead to severe degradation and the phoneme information is not fully explored in enhancement. 

Inspired from the operation in batch normalization, we propose a phoneme-based distribution regularization (PbDr) to integrate recognition information into the speech enhancement network in a conditional manner to achieve better speech quality. Different phonemes often lead to different feature distributions in frequency. Based on the phoneme probability vector from a robust phoneme classification network, we learn a modulation parameter pair, i.e. scale and bias, to regularize the features of the speech enhancement network. In this way, features are regularized into phoneme-related distributions. Experiments show that the proposed PbDr module could not only improve the perceptual quality but also the recognition accuracy of an ASR system on the enhanced speech.    

\begin{figure*}[htb]

\begin{minipage}[b]{1.0\linewidth}
  \centering
  \centerline{\includegraphics[width=18cm]{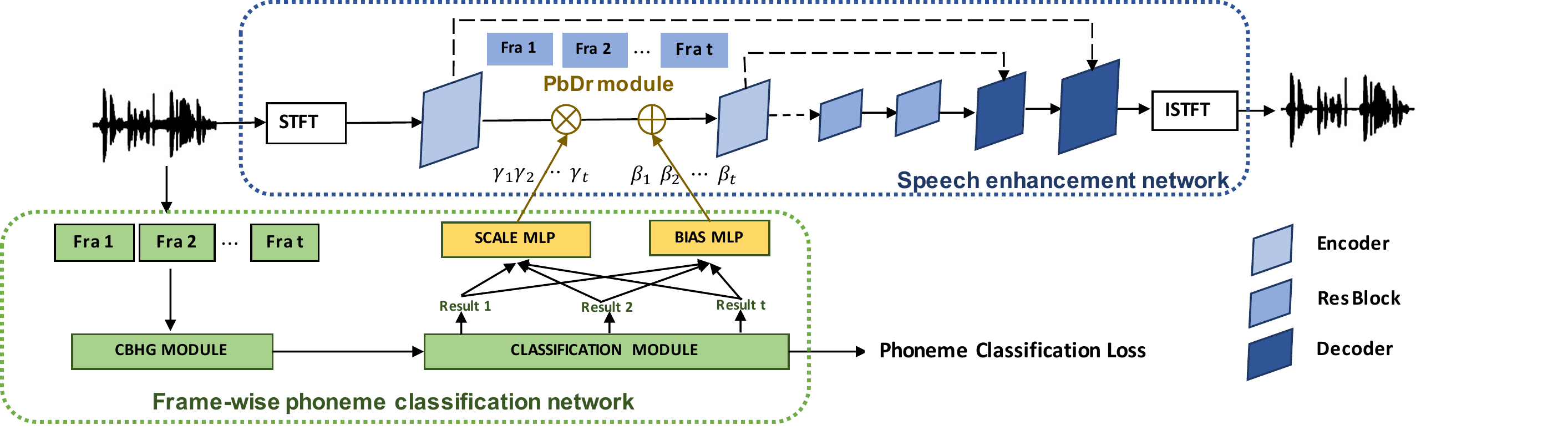}}
  \vspace{-0.2cm}
\end{minipage}
 \vspace{-0.5cm}
\caption{The architecture of PbDrNet. The PbDrNet consists of two parts, i.e. the frame-wise phoneme classification network and a baseline speech enhancement network. The PbDr module learns distribution modulation parameter pairs based on the phoneme classification vectors and then maps features in speech enhancement network to the phoneme-related distributions.}
\label{fig:res}
\end{figure*}
\section{Related Work}
\label{sec:format}
Recent methods propose to apply recurrent neural networks, convolutional neural networks, and generative adversarial networks to address the speech enhancement problem \cite{williamson2015complex, erdogan2015phase, pandey2018adversarial}. Comparatively, gated residual networks (GRN) \cite{tan2018gated} and convolutional recurrent networks (CRN) \cite{tan2018convolutional} with an encoder-decoder structure achieve better generalization for unknown situations. However, these methods only regularize the similarity of the enhanced speech and the target, without paying attention to the semantic information of the corrupted speech.

Recent studies \cite{wang2016phoneme,schulze2020joint,chazan2016phoneme} attempt to introduce phoneme information to a speech enhancement network. \cite{wang2016phoneme} proposes a phoneme-specific network for speech enhancement. However, it needs a separate model for each phoneme and wrong phoneme predictions will lead to severe degradation in enhanced speech. \cite{schulze2020joint} aims to jointly achieve phoneme alignment and text-informed speech separation. However, it needs to feed known text information as side information. In this paper, we propose to learn distribution modulation parameters from phoneme classification vectors and utilize them to map enhanced features to phoneme-related distributions. 

\section{Approach}
A corrupted speech $c$ is a mixture of clean speech $s$ and a noise signal $n$. The goal of a speech enhancement network is to reconstruct the clean signal $\hat{s}$ from $c$ with good perceptual quality. In this paper, we propose the phoneme-based distribution regularization network (PbDrNet), an integration of the PbDr module into a baseline speech enhancement network.

\subsection{Overview of PbDrNet}
As shown in Figure 1, the proposed PbDrNet consists of two parts, i.e. the phoneme classification network in green and the speech enhancement network in blue. For each frame, we get a probability vector from the phoneme classification network, showing the probabilities of different phoneme categories. The probability vector is then fed into the PbDr module to obtain a modulation parameter pair, i.e. scale and bias. Inspired from the distribution regularization in batch normalization, the modulation parameters are multiplied and added to a target feature to regularize the distributions of features in speech enhancement network. Due to the importance of semantic information in the network, this modulation is performed on early encoder layers to guide the enhancement process in the network. In this way, we effectively integrate the frame-wise phoneme prior to speech enhancement network.

\subsection{Speech Enhancement Network}

The speech enhancement network takes complex time-frequency (T-F) spectrum $C$ as input. It consists of an encoder, four residual blocks and a decoder. There are three convolutional layers followed by max-pooling layers in encoder. Each residual block includes two convolutional layers with a shortcut connection. The decoder includes three deconvolution layers to restore the resolution of original speech. Furthermore, there are skip connections between encoder and decoder to recover back information loss. Following \cite{choi2018phase}, the decoder outputs an amplitude gain and the phase, which are used to recover the T-F spectrum of the estimated speech $\hat{S}$. After an inverse STFT, the estimated clean speech in the time domain $\hat{s}$ is generated.

\subsection{Frame-wise Phoneme Classification Network}
For robustness, the MFCC feature is selected as the input to the phoneme classification network. This network includes a CBHG\cite{wang2017tacotron} and a classification module, the former of which extracts comprehensive features and the latter classification module outputs the predicted phoneme category. The CBHG module consists of several 1-D convolutional filters, highway networks and a bidirectional gated recurrent unit (GRU). These 1-D convolutional filters explore local and contextual information while the highway networks and the GRU extract high-level information. For phoneme classification, the following cross-entropy loss is used for optimization, i.e.
\begin{equation}
L_{phoneme}=-\sum\limits_{k=1}^{K}\sum\limits_{t=1}^{T_k} log(p_{c_k^t, y_k^t}), 
\end{equation}
where K denotes the number of audios in a batch, and $T_k$ is the number of frames in the $k$-th audio. $c_k^t$ and $y_k^t$ represent the noisy speech and the true phoneme label for $t$-th frame of the $k$-th audio, respectively. $p_{c_k^t, y_k^t}$ is the predicted probability for $c_k^t$ is of class $y_k^t$.

\begin{figure}[t]

\begin{minipage}[b]{1.0\linewidth}
  \centering
  \centerline{\includegraphics[width=9cm]{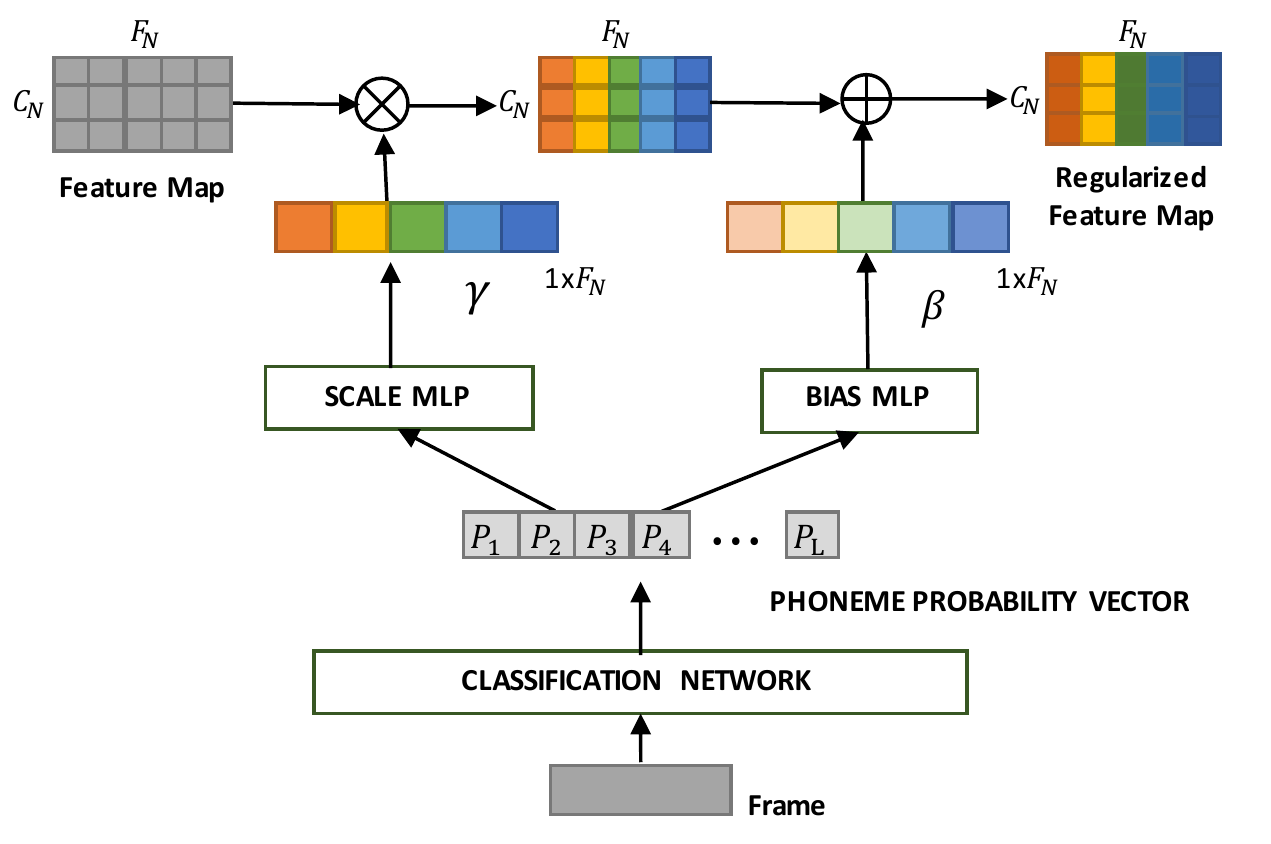}}
%
\end{minipage}

\caption{The architecture of the PbDr module.}
\label{fig:res}
\end{figure}

\subsection{PbDr Module}
Different phoneme categories always lead to different distributions in frequency. Inspired from the operation in batch normalization, as shown in Figure 2, the PbDr module learns distribution modulation parameter pairs based on the classification probability vector, which are multiplied and added to regularize the distributions of features in enhancement network.

The predicted hard label for noisy audio is not always accurate. To obtain stronger robustness, we utilize the probability vector $\bm{P}\in R^{1\times L}$ generated by the classification network for each frame to assist the speech enhancement network, where $L$ denotes the number of phoneme categories. Based on the prior phoneme probability vector $\bm{P}$, the PbDr module learns a mapping function $\mathcal{M}$ to obtain a modulation parameter pair $(\bm{\gamma},\bm{\beta)}$, which can be formulated as
\begin{equation}
 \vspace{-0.25cm}
(\bm{\gamma},\bm{\beta}) = \mathcal{M}(\bm{P}),
\end{equation}where $\bm{\gamma},\bm{\beta}\in R^{1\times F_N}$ represent the scale and bias modulation parameter, respectively, and $F_N$ is the number of frequency bins of feature maps in the $N$-th layer. As the frequency dimension is maintained in our network with reduced resolution in encoder, we perform both a frame-wise and a coarse frequency-wise condition to the network. Additionally, $\mathcal{M}$ is applied by two multi-layer perceptrons (MLP). 

With the learned modulation parameters $(\bm{\gamma},\bm{\beta)}$, an intermediate feature in the enhancement network is adaptively mapped to phoneme-related distributions through an affine transformation given by
\begin{equation}
PbDr(\bm{F}|\bm{\gamma},\bm{\beta})=\bm{F}\otimes\bm{\gamma} \oplus \bm{\beta},
 \vspace{-0.25cm}
\end{equation}
where $\bm{F}\in R^{F_N\times C_N}$ denotes the feature map for each frame, and $C_N$ is the number of channels for feature maps in the $N$-th layer. $\otimes,\oplus$ represent the element-wise multiplication and addition in frequency dimension, and we apply the same modulation parameter to all channels.

Furthermore, where to integrate the phoneme information into the enhancement network is also crucial. Due to the importance of semantic information in the network, the modulation is performed on early encoder layers to guide the enhancement process in the network. Additionally, as the resolution of feature maps changes in different layers of the encoder, the condition can be applied in different scales. Adding phoneme-based conditional information to a larger scale means we pay more attention to the high-level information of the speech. On the contrary, if we add conditional information to a small scale, it means we pay more attention to details in the speech. In the experiments, we will further investigate these issues.

\subsection{Loss Function}
We apply L1 Loss in the frequency domain to match human perceptions. To keep STFT consistency \cite{wisdom2019differentiable}, the enhanced speech spectrum is firstly transformed to time domain and then to the frequency domain to calculate the loss. In our PbDrNet, the two sub-networks are jointed optimized based on the combined loss as below
\begin{equation}
L_{all} = \sum\limits_{k=1}^{K}|\hat{S_k}-S_k| + \lambda L_{phoneme}, 
 \vspace{-0.25cm}
\end{equation}
where $S_k$ and $\hat{S}_k$ represent the target and estimated complex spectrum for speech s. $\lambda$ is the weight for phoneme classification loss, which is set to 1 in our experiment.
\label{sec:pagestyle}

 \vspace{-0.25cm}
\section{experiments}
\label{sec:typestyle}

\subsection{Dataset and Setting}
We utilize speech material from a 100-hour subset of the public Librispeech dataset\cite{panayotov2015librispeech}. The Librispeech dataset is mixed up with noise in AudioSet dataset to generate noisy speech. For training, we utilize 21855 utterances in the 100-hour subset. For testing, 500 utterances are selected. Each clean utterance is mixed up with two noise audios for training and one noise for testing, with a SNR uniformly drawn from [-5,25] dB. The STFT is computed with a window length of 20 ms and a hop length of 10 ms. The Hamming window is used for overlapped frames. Following \cite{liu2020mockingjay}, we obtain the force-aligned phoneme labels by Montreal Forced Aligner \cite{mcauliffe2017montreal}, which has 72 phoneme classes. The input of the phoneme classification network are 13-dimension MFCC features computed with a window size of 40 ms, and a hop length of 10 ms. 

During training, we use a batch size of 32 and the ADAM optimizer for all schemes. The learning rates for the enhancement and the phoneme classification network are set to 0.0002 and 0.001, respectively. The training is stopped after 250 epochs.

\begin{table}[t]
  \centering
  \fontsize{8}{10}\selectfont
  \setlength\abovedisplayskip{1pt}
\setlength\belowdisplayskip{1pt}
  \label{tab:performance_comparison}
      \caption{Comparison of enhancement quality evaluation results among different methods.}
\begin{tabular}{|c|c|c|c|c|c|}
\hline
 \  & CSIG  &CBAK & COVL &PESQ &SSNR\\  
\hline
 NOISY &3.35 & 2.94 & 2.59 &1.86&10.28 \\  
 \hline
 BASELINE &3.96 & 3.72 & 3.36 &2.72 &15.12 \\  
 \hline
 CASCADE &3.91 &3.74 & 3.34 &2.74 &15.16 \\  
 \hline
 PbDrNet &4.06 &3.82 &3.48 &2.85 &\bf{15.44} \\  
 \hline
 E-PbDrNet &\bf{4.14} &\bf{3.83} &\bf{3.54} &\bf{2.89} &15.31 \\  
 \hline
\end{tabular}
\label{compare}
\end {table}


\begin{table}[t]
  \centering
  \fontsize{8}{10}\selectfont
  \setlength\abovedisplayskip{1pt}
\setlength\belowdisplayskip{1pt}
  \label{tab:performance_comparison}
      \caption{Ablation studies for PbDrNet.}
\begin{tabular}{|c|c|c|c|c|c|}
\hline
 \  & CSIG  &CBAK & COVL &PESQ &SSNR \\  
\hline
 BASELINE &3.96 & 3.72 & 3.36 &2.72 &15.12 \\  
 \hline
  CONCAT &3.82 &3.75&3.29 & 2.74 &15.41 \\  
 \hline
 PbDrNet\_1 &3.88 &3.81 &3.37 &2.81 &\bf{15.66}\\  
 \hline
 PbDrNet\_2 &\bf{4.06} &\bf{3.82} &\bf{3.48} &\bf{2.85} &15.44\\  
 \hline
\end{tabular}
 \vspace{-0.5cm}
    \label{compare}
\end {table}

\subsection{Speech Enhancement Results}
Five commonly used evaluation metrics are utilized here to evaluate the quality of enhanced speech, i.e. CSIG (mean opinion score predictor of signal distortion), CBAK (mean opinion score predictor of background-noise intrusiveness), COVL (mean opinion score predictor of overall signal quality), PESQ (perceptual evaluation of speech quality) and SSNR (segmental SNR). We first compare PbDrNet with the following three methods.

\noindent \textsl{BASELINE}: only apply the speech enhancement network.

\noindent \textsl{CASCADE}: cascade a pretrained phoneme classification network to the speech enhancement network, and finetune them together. This is a straightforward way to regularize the recognizability of the enhanced speech. 

\noindent \textsl{E-PbDrNet}: first utilize one enhancement network to get an enhanced speech and then learn the phoneme information from the enhanced signal rather than noisy to modulate features of another enhancement network. This is to show whether more accurate phoneme information could lead to better speech enhancement. 

As shown in Table 1, the CASCADE method performs comparatively with BASELINE, showing that it is not effective as there might be gradient vanishing problem to cascade two deep networks for optimization. Our PbDrNet more effectively regularizes the distribution of the enhanced speech and achieves obvious advantages over BASELINE in all the five metrics. There are 0.18, 0.11, 0.18, 0.17 and 0.19dB gains in CSIG, CBAK, COVL, PESQ and SSNR, respectively.

For E-PbDrNet, as it learns phoneme information on the enhanced speech rather than noisy, it achieves a more accurate classification with the top1 accuracy increasing from 77.72\% to 80.49\% and top3 from 95.50\% to 96.81\%. As shown in Table 1, based on the more precise classification results, E-PbDrNet obtains better results in CSIG, CBAK, COVL and PESQ and slightly worse result in SSNR compared with PbDrNet. This shows that iteratively learning the enhancement and recognition networks can help to promote each other and arrive at a good state.

We further investigate how and where to integrate the phoneme information into the enhancement network. In Table 2, CONCAT denotes a direct concatenation of features learned from the phoneme classification network with that of the speech enhancement network. The little gap with BASELINE shows that this simple concatenation is poor to incorporate semantic information into speech enhancement. As the selection of scale to incorporate the condition vector in PbDrNet is important, we further compare PbDrNet\_1 and PbDrNet\_2, which add conditional information in the first and second convolutional layers in the encoder, respectively. Information fusion on a small scale, i.e. PbDrNet\_1, means we pay more attention to details in the speech; therefore it achieves a higher SSNR. On the contrary, adding conditional information on a larger scale, i.e. PbDrNet\_2, means we pay more attention to higher level information; thus it gets a higher number in CSIG, COVL and PESQ. Unless otherwise specified, PbDrNet\_2 is used in PbDrNet in all experiments. 

\begin{table}[t]
  \centering
  \fontsize{8}{10}\selectfont
  \setlength\abovedisplayskip{1pt}
\setlength\belowdisplayskip{1pt}
  \label{tab:performance_comparison}
      \caption{Comparison of speech recognition accuracy(\%) among different methods.}
\begin{tabular}{|c|c|c|}
\hline
 \  & WER  &CER  \\  
\hline
 NOISY &30.57 & 16.28  \\  
 \hline
 BASELINE &15.60 & 7.30 \\  
 \hline
 CASCADE &15.54 &7.23  \\  
 \hline
 PbDrNet &13.64 &6.18\\  
 \hline
 E-PbDrNet &\bf{12.92} &\bf{5.96}\\  
 \hline
\end{tabular}
 \vspace{-0.5cm}
    \label{compare}
\end {table}
\subsection{Speech Recognition Results}

%
In this section, we evaluate the recognizability of the enhanced speech obtained by all methods mentioned in Table 1. We utilize the publicly released speech recognition model by DeepSpeech2 \cite{amodei2016deep} for this evaluation. Word error rate (WER) and character error rate (CER) are selected as evaluation metrics. As shown in Table 3, the speech recognition results align with the enhancement results. The CASCADE method slightly increases the recognition accuracy while the proposed PbDrNet significantly reduces the WER and CER by 12.6\% and 15.3\% compared with BASELINE, respectively. The E-CPbDrNet achieves the best result with 17.2\% and 18.4\% reduction in WER and CER, showing that more precise phoneme information can better improve the enhanced speech in recognizability.

 \vspace{-0.25cm}
\section{Conclusion}
We propose the phoneme-based distribution regularization to utilize speech recognition information to help speech enhancement. Experiments demonstrate its effectiveness in not only improving perceived quality but also the recognizability of the enhanced speech for an ASR system. Without restricted by the network in PbDrNet, the PbDr module could be readily employed to other speech enhancement networks with different losses as well.

\bibliographystyle{IEEEbib}
\bibliography{reference.bib}

\end{document}